\documentclass[12pt,technote, onecolumn, draftclsnofoot]{IEEEtran}
\usepackage{cite}
\usepackage[justification=centering]{caption}
\usepackage{amsmath,amssymb,amsfonts}
\usepackage{setspace}
\usepackage{algorithmic}
\usepackage{graphicx}
\usepackage{subfigure}
\usepackage{textcomp}
\usepackage{xcolor}
\usepackage{times,amssymb,amsmath,amsfonts,float,nicefrac,color,subfigure,bm}
\usepackage[ruled,vlined]{algorithm2e}
\usepackage{tikz}
\usepackage{bbding}
\usepackage{diagbox}
\usepackage[dvips]{epsfig}
\usepackage{verbatim}
\usepackage{pgfplots}
\usepackage{multirow}
\usepackage{arydshln}
\usepackage{verbatim}
\usepackage{pgf}

\newtheorem{theorem}{Theorem}

\newtheorem{definition}{Definition}
\newtheorem{remark}{Remark}

\newtheorem{lemma}{Lemma}
\newtheorem{proposition}{Proposition}

\def\BibTeX{{\rm B\kern-.05em{\sc i\kern-.025em b}\kern-.08em
    T\kern-.1667em\lower.7ex\hbox{E}\kern-.125emX}}

\begin{document}

\title{Bounds and Constructions of Singleton-Optimal Locally Repairable Codes with Small Localities}
\author{Weijun Fang, Bin Chen, Shu-Tao Xia, Fang-Wei Fu
\IEEEcompsocitemizethanks{\IEEEcompsocthanksitem Weijun Fang is with School of Cyber Science and Technology, Shandong University, Qingdao, 266237, P.R. China. (Email: fwj@sdu.edu.cn.)
}
\IEEEcompsocitemizethanks{\IEEEcompsocthanksitem Bin Chen is with Department of Computer Science and Technology, Harbin Institute of Technology, Shenzhen, 518055, P.R. China. (Email: chenbin2021@hit.edu.cn)}

\IEEEcompsocitemizethanks{\IEEEcompsocthanksitem Shu-Tao Xia are with Tsinghua Shenzhen International Graduate School, Tsinghua University, Shenzhen 518055, P.R. China  (Email:  xiast@sz.tsinghua.edu.cn.)
}

\IEEEcompsocitemizethanks{\IEEEcompsocthanksitem Fang-Wei Fu is with Chern Institute of Mathematics and LPMC, Nankai University, Tianjin 300071, P.R. China. (Email: fwfu@nankai.edu.cn.)}
\thanks{$^\dag$This research is supported in part by National Key R$\&$D Program of China under Grant Nos. 2021YFA1001000 and 2018YFA0704703, the National Natural Science Foundation of China under Grants 62171248, 61971243, the Fundamental Research Funds for the Central Universities, Nankai University, the Natural Science Foundation of Tianjin (20JCZDJC00610).}}



\maketitle

\begin{abstract}
Constructions of optimal locally repairable codes (LRCs) achieving Singleton-type bound have been exhaustively investigated in recent years. In this paper, we consider new bounds and constructions of Singleton-optimal LRCs with minmum distance $d=6$, locality $r=3$ and minimum distance $d=7$ and locality $r=2$, respectively. Firstly, we establish  equivalent connections between the existence of these two families of LRCs and the existence of some subsets of lines in the projective space with certain properties. Then, we employ the line-point incidence matrix and Johnson bounds for
constant weight codes to derive new improved bounds on the code length, which are tighter than known results. Finally, by using some techniques of finite field and finite geometry, we give some new constructions of Singleton-optimal LRCs, which have larger length than previous ones.

\end{abstract}


\section{Introduction}

In a large-scale distributed storage system (DSS), node failures  frequently occur and it is critical to recover a failed node in time.
Due to the large storage overhead of replication strategy, erasure codes were introduced to increase the storage efficiency and reduce the cost during repairing a failed node. To  minimize the number of storage nodes involved in repairing a failed node,  Gopalan \emph{et al.} \cite{GHSY12} introduced locally repairable codes to recover a failed node by accessing few available nodes. Suppose $\mathcal{C}$ is an $[n, k, d]_q$ linear code.
The $i$-th symbol of $C$ has \emph{locality} $r$ if  there exists a subset $R_i\subset [n]=\{1,2, \cdots, n\}$ with $i \in R_i$ and $|R_{i}| \leq r+1$ such that the $i$-th symbol $c_i$ can be represented as a linear combination of $\{c_j\}_{j\in R_i\setminus\{i\}}$. The subset $R_i$ is called the \emph{repair group} of the $i$-th symbol. $C$ is called a \emph{locally repairable code} (LRC) with locality $r$ if each symbol of $C$ has locality $r$.  We denote such a code as an $(n, k, d ;r)$-LRC. An LRC is saied to have \emph{disjoint local repair groups} if the set of coordinates $[n]$ is exactly the disjoint union of some repair groups of size $r+1$, i.e., $[n]=\bigsqcup_i R_i,$ and $|R_{i}|=r+1$. In this paper, we mainly focus on LRCs with disjoint local repair groups. The reason is that LRCs with disjoint local repair groups are widely adopted in constructing LRCs and have more advantages in performance.  Any $q$-ary $(n, k, d ;r)$-LRCs have to satisfy the well-known Singleton-type bound in \cite{GHSY12},
\begin{equation}
\label{singleton}
d\leq n-k-\left\lceil \frac{k}{r}\right\rceil+2.
\end{equation}
 A $q$-ary $[n, k, d]$ linear code $\mathcal{C}$ with locality $r$ is called a \emph{Singleton-optimal} $(n, k, d; r)$-LRC if it achieves the bound \eqref{singleton} with equality. Constructing Singleton-optimal LRCs have been exhaustively investigated in the literature (\cite{GHSY12,SDYL14,PKLK12,GXY19,FCXF20-2,CFXHF20,Hao,ternary,TB14, CXHF18, CFXF19,LXY19}). When the locality $r=k$, the Singleton-type bound reduces to the Classical Singleton bound. Thus Singleton-optimal LRCs are analog to MDS codes. For any given $d$ and $r$, a Singleton-optimal LRC with
longer length $n$ would have larger code rate $k/n$. Thus it is interesting to construct Singleton-optimal LRCs with large length and deriving an
upper bound on the maximal code length.

In \cite{TB14}, Tamo and Barg first gave a breakthrough construction of Singleton-optimal LRCs via subcodes of Reed-Solomon codes whose code length can go up to the alphabet size. Chen \emph{et al.} \cite{CXHF18,CFXF19} completely determined the $q$-ary Singleton-optimal LRCs of length $n \mid (q+1)$ for all possible parameters via cyclic and constacyclic codes. In \cite{JMX19}, Jin \emph{et al.} also constructed a family of $q$-ary Singleton-optimal $r$-LRCs with length up to $q + 1$ by using the automorphism group of rational function fields. For some particular $r=2,3,5,7,11$ or $23$, Singleton-optimal LRCs with lengths up to $q+2\sqrt{q}$ are constructed via elliptic curves \cite{LMX19}. When the minimum distance $d=3$ or $4$, Luo \emph{et al.} \cite{LXY19} proposed a construction of cyclic Singleton-optimal LRCs whose length can be arbitrarily large.   Analog to the well-known MDS conjecture, the problem of the maximal code length of Singleton-optimal LRCs has attracted the attention of many scholars. For minimum distance $d \geq 5$, Guruswami \emph{et al.} \cite{GXY19} proved that the code length of a $q$-ary Singleton-optimal LRC must be at most roughly $O(dq^{3})$. Jin \cite{J19} presented an explicit construction of $q$-ary optimal LRCs of length $\Omega_{r}(q^{2})$ via binary constant weight codes. In \cite{XY18}, Xing and Yuan generalized Jin's results and presented a construction of $q$-ary Singleton-optimal LRCs for general $d \geq 7$ via hypergraph theory. Recently, Cai et al. \cite{CMST20} improved these results and generalized them to the $(r, \delta)$-LRCs. In our prior work \cite{FCXF20-2,CFXHF20}, we considered the case of  $d=6$ and $r=2$ and provided some new constructions and bounds of Singleton-optimal LRCs with larger code length. We summarize some known constructions of $(n,k,d;r)$-Singleton-optimal LRCs in Table I.
\begin{table}[]
    \centering
        \caption{Some known constructions of Singleton-optimal $(n,k,d;r)$-LRCs}
 \begin{tabular}{|c|c|c|c|}
 \hline
    Length $n$ &  Minimum distance  $d$ & Locality $r$ & References\\
     \hline
     $n=q$ & $d \leq n$ & $r \leq k$ & \cite{TB14}\\
       \hline
      $n=q+1$ & $d \leq n$ & $r \leq k$& \cite{CXHF18,CFXF19,JMX19}\\
      \hline
       $n=q+2\sqrt{q}$  & $(r+1) \mid d$ & $r=2,3,5,7,11,23$& \cite{LMX19}\\
         \hline
      Unbounded   & $d=3,4$ & $r \geq 2, r+1 \mid q+1$ & \cite{LXY19}\\
      \hline
     $n=3(2q-4)$  & $d=6$ & $r=2$& \cite{FCXF20-2,CFXHF20} \\
          \hline
        $n=\Omega_{r}(q^{2})$  & $d=5, 6$ & $r \geq 4$, $r$ is a prime power& \cite{J19,XY18}\\
           \hline
           $n=q^{2-o(1)}$  & $d=7, 8$ & $r \geq d-2$& \cite{XY18}\\
            \hline
           $n=q^{3/2-o(1)}$  & $d=9, 10$ & $r \geq d-2$& \cite{XY18}\\
           \hline
           $n=\Omega_{q,r}(q(q\log q)^{\frac{1}{\lfloor\frac{d-3}{2}\rfloor}})$  & $d \geq 11$ & $r \geq d-2$& \cite{XY18}\\
           \hline
 \end{tabular}

    \label{con}
\end{table}

In this paper, we consider new bounds and constructions of Singleton-optimal LRCs with minmum distance $d=6$, locality $r=3$ and minimum distance $d=7$ and locality $r=2$, respectively. By using the approach of parity-check matrix, we establish  equivalent connections between the existence of these two families of LRCs and the existence of some subsets of lines in projective geometry with special structures. New improved bounds  and optimal constructions of LRCs with large length are then obtained. We summarize some known  bounds on the code length of $(n, k, d;r)$ Singleton-optimal LRCs for $d=6, r=3$ and $d=7, r=2$ in Table II and list our main results as follows.

\begin{table}[]
    \centering
        \caption{Some known bounds on the code length of Singleton-optimal $(n,k,d;r)$-LRCs for $d=6, r=3$ and $d=7, r=2$}
 \begin{tabular}{|c|c|c|c|}
 \hline
    Minimum distance  $d$  & Locality $r$  & Length $n$ & References\\
     \hline
     $d=6$ & $r=3$ & $n \leq \frac{4q^4}{3(q-1)}$ & \cite{GXY19}\\
       \hline
      $d=6$ & $r=3$ & $n \leq 4\lfloor\frac{q^2+q+1}{6}\rfloor$& \cite{CFXHF20}\\
      \hline
       $d=7$  & $r=2$ & $n \leq \frac{3}{2}(\frac{q^3}{q-1}+1)$& \cite{GXY19}\\
         \hline
      $d=7$  & $r=2$ & $n \leq 3\lfloor\frac{(q^2+1)(q+1)}{3}\rfloor$ & \cite{CFXHF20}\\
          \hline
 \end{tabular}

    \label{con}
\end{table}

\begin{itemize}
\item We provide sufficient and necessary conditions on the existence of Singleton-optimal $(n,k,d;r)$-LRCs for $d=6, r=3$ and $d=7, r=2$ with disjoint repair groups, respectively, which establish equivalent connections with some subsets of lines in the projective space with certain properties (see Theorems \ref{d6r3} and \ref{thm8});
\item For any $q$-ary Singleton-optimal $(n,k,d=6;r=3)$-LRCs with disjoint repair groups, we prove that $n =O(q^{1.5})$ (see Theorem \ref{thm4});
\item We give an explicit construction of $q$-ary Singleton-optimal $(n,k,d=6;r=3)$-LRCs with $n=q+4$ and $q=2^m$ (see Theorem \ref{thm7});
\item For any $q$-ary Singleton-optimal $(n,k,d=7;r=2)$-LRCs with disjoint repair groups, we prove that $n \leq 3\lfloor\frac{q^2+q+3}{3}\rfloor$ (see Theorem \ref{thm9});
\item We show the existence of $q$-ary Singleton-optimal $(n,k,d=7;r=2)$-LRCs with $n \approx \sqrt{2}q$ (see Theorems \ref{thm10} and \ref{thm11}).
\end{itemize}

According to Tabels I and II, our new bounds on the code length improve the previous results and our new constructions of Singleton-optimal LRCs have larger length than previous constructions.

The rest of this paper is organized as follows. In Section II, we introduce some basic results on LRCs, finite geometry and the Johnson bound for constant weight codes. In Section III, we consider the new constructions and bounds of Singleton-optimal $(n,k,d=6;r=3)$-LRCs. In Section IV, we consider the new constructions and bounds of Singleton-optimal $(n,k,d=7;r=2)$-LRCs. We conclude this paper in Section V.

\section{Preliminaries}
In this section, we introduce some basic notation and results about LRCs, finite geometry and constant weight codes.

\subsection{LRCs with Disjoint Repair Groups}
Suppose $q$ is prime power and $\mathbb{F}_{q}$ is a finite field with $q$ elements. Denote $[n] \triangleq \{1,2,\dots, n\}$. For a vector $\bm{v}=(v_{1}, v_{2}, \cdots, v_{n}) \in \mathbb{F}_{q}^{n}$, the support of $\bm{v}$ is defined as $\text{supp}(\bm{v})\triangleq \{i \in [n]: v_{i} \neq 0\}$.

First, we recall the parity-check matrix approach to LRCs proposed in \cite{HXSCFY20,CFXHF20}. Let $C$ be an $[n, k, d]_{q}$-linear code and $(r+1) \mid n$. Then $C$ is an LRC with locality $r$ if and only if for each $i \in [n]$, there exists a codeword $\bm{h}_{i}$ of the dual code $C^{\perp}$ such that $i \in \text{supp}(\bm{h}_{i})$ and $wt(\bm{h}_{i}) \leq r+1$. Throughout this paper, we assume that the LRCs have \emph{disjoint local repair groups}. Thus there exist $\ell\triangleq \frac{n}{r+1}$ vectors $\bm{h}_{1}, \bm{h}_{2}, \ldots, \bm{h}_{\ell}$ of $C^{\perp}$ which are called \emph{locality vectors}, such that $wt(\bm{h}_{i})=r+1$ and $\text{supp}(\bm{h}_{i}) \cap \text{supp}(\bm{h}_{j})=\emptyset$ for any $1 \leq i \neq j \leq \ell$. Since two equivalent codes have the same code length, dimension, minimum distance and locality. Thus an LRC with disjoint local repair groups has an equivalent parity-check matrix $H$ as the following form:
 \begin{equation}\label{5}
 \setlength{\arraycolsep}{4pt}
\left(
  \begin{array}{cccc|cccc|c|cccc}
    1 & 1 &  \!\!\!\dots \!\!\! & 1 & 0 & 0 & \!\!\! \dots \!\!\! & 0 &  \!\!\!\dots \!\!\! & 0 & 0 & \!\!\!\dots\!\!\! & 0\\
    0 & 0 &  \!\!\!\dots \!\!\! & 0 & 1 & 1 &  \!\!\!\dots \!\!\! & 1 &  \!\!\!\dots \!\!\! & 0 & 0 &  \!\!\!\dots  \!\!\!& 0\\
  \vdots & \vdots & \vdots & \vdots & \vdots & \vdots & \vdots & \vdots &  \!\!\!\dots \!\!\! & \vdots & \vdots &  \vdots & \vdots \\
    0 & 0 &  \!\!\!\dots \!\!\! & 0 & 0 & 0 &  \!\!\!\dots \!\!\! & 0 & \! \dots \! & 1 & 1 &  \!\!\!\dots  \!\!\!& 1 \\
    \hline
    \mathbf{0} &\bm{v}_{1}^{(1)} & \!\!\!\dots \!\!\! &\bm{v}_{r}^{(1)} & \mathbf{0} & \bm{v}_{1}^{(2)} &\!\!\dots  & \!\!\!\bm{v}_{r}^{(2)} & \dots & \mathbf{0} & \bm{v}_{1}^{(\ell)} &\!\!\dots\!\! & \bm{v}_{r}^{(\ell)}
  \end{array}
\right),
\end{equation}
where the upper part of $H$ contains $\ell$ locality rows and the lower part of $H$  contains $u\triangleq n-k-\ell$ rows to ensure that $rank(H)=n-k$. The bold-type letters $\bm{v}_{i}^{(j)}$, $i\in[\ell]$, $j\in [r]$ are column vectors in $\mathbb{F}_q^{u}$, while ${\bf 0}$ represents the all-zero column vector in $\mathbb{F}_q^{u}$.

\subsection{Singleton-Optimal LRCs}
The Singleton-optimal LRCs could be equivalently characterized by an alternative equality as follows.
\begin{lemma}(\cite[Lemma II.2]{GXY19})\label{lem1}
Let $n,k,d,r$ be positive integers with $(r+1)\mid n$. If the Singleton-like bound  is achieved, then
\begin{equation}\label{3}
n-k=\ell+d-2-\left\lfloor\frac{d-2}{r+1}\right\rfloor.
\end{equation}
Conversely, if $d-2\not\equiv r(\textnormal{mod~} r+1)$ and the equality \eqref{3} is satisfied, then  the Singleton-like bound (\ref{singleton}) is achieved.
\end{lemma}
\begin{remark}\label{rem1}
If $d - 2 \equiv r (\textnormal{mod~}r + 1)$, by
[7, Corollary 10] one cannot achieve the Singleton-type bound
with equality and one must have $d\leq n-k-\left\lceil \frac{k}{r}\right\rceil+1.$
Therefore in this paper we refer the equality \eqref{3} as the
Singleton-type bound of LRCs.
\end{remark}

\subsection{Finite Geometry}
In this subsection, we introduce some basic notion and results of geometry over finite fields, which will be used in Sections III and IV. And we refer the readers to \cite{ES16, MS77, C06, CD06, H79} for more details about finite geometry.

Let $\mathbb{F}_{q}^{N+1}$ be the $(N+1)$-dimensional vector space over $\mathbb{F}_{q}$ with origin $\bm{0}$. We consider the equivalence relation on the vectors of $\mathbb{F}_{q}^{N+1}\setminus \{\bm{0}\}$ whose equivalence classes are the one-dimensional subspaces of  $\mathbb{F}_{q}^{N+1}$ without the origin. Precisely, for two nonzero vectors $\bm{u}=(u_{1}, u_{2}, \dots, u_{N+1}),  \bm{v}=(v_{1}, v_{2}, \dots, v_{N+1}) \in \mathbb{F}_{q}^{N+1}\setminus \{\bm{0}\}$, we say $\bm{u}$ is equivalent to $\bm{v}$ if there exists $t \in \mathbb{F}_{q}^{*}$ such that $\bm u=t \bm v$.

\begin{definition}[\cite{H79}]
The $N$-\emph{dimensional projective space} over $\mathbb{F}_{q}$ is defined as the set of equivalence classes and is denoted by $PG(N,q)$.  The elements of $PG(N,q)$ are called \emph{points}. In particular, when $N=2$, $PG(N,q)$ is called the \emph{projective plane}.
\end{definition}

If  the point $P(\bm{u})$ is the equivalence class of the vector $\bm{u} \in \mathbb{F}_{q}^{N+1}\setminus \{\bm{0}\}$, we say that $\bm{u}$ is a vector representing  $P(\bm{u})$. We say that the points $P(\bm{u}_{1}), P(\bm{u}_{2}), \dots, P(\bm{u}_{s})$ are collinear if the vectors $\bm{u}_{1}, \bm{u}_{2}, \dots, \bm{u}_{s}$ are linearly dependent. \emph{A subspace of dimension} $m$, or an $m$-\emph{subspace} of $PG(N,q)$ is a set of points all of whose representing vectors form (together with the origin) a subspace of dimension $m+1$ of  $\mathbb{F}_{q}^{N+1}$. A subspace of dimension zero has already called \emph{a point}. Subspaces of dimensions one and two are called a \emph{line} and a \emph{plane}, respectively.

Some basic properties of  $PG(N,q)$ are summarized in the following.
\begin{lemma}[\cite{H79}]\label{lem2}
 \begin{description}
   \item[(i)] Each line in $PG(N,q)$ has $q+1$ points;
   \item[(ii)] The number of lines in $PG(N,q)$ throughing a fixed point is $\frac{q^N-1}{q-1}$;
   \item[(iii)] Any two distinct lines in the projective plane $PG(2,q)$ meet at exactly one point;
   \item[(iv)] The number of points in the projective plane $PG(2,q)$ is $q^2+q+1.$
 \end{description}
\end{lemma}

The following is the well-known principle of duality in the projective plane $PG(2,q)$.
\begin{theorem}\cite[Theorem 2.1,The Principle of Duality]{C06}\label{dual}
If $T$ is a theorem valid in the projective plane $PG(2,q)$, and $T'$ is the statement obtained from $T$ by making the following changes:
\[point \leftrightarrow line,\]
\[collinear \leftrightarrow concurrent,\]
\[join \leftrightarrow intersection,\]
and whatever grammatical adjustments that are necessary, then $T'$ (called the
Dual Theorem) is a valid theorem in the $PG(2,q)$.
\end{theorem}

Finally, we introduce the definitions of spreads and sunflowers in projective space, respectively.
\begin{definition} [Spreads in the projective space \cite{CD06}]
Suppose $t \leq N$. A $t$-spread in a $N$-dimensional projective space $PG(N,q)$ is a set $S$ of $t$-dimensional subspaces such that any point of $PG(N,q)$ is on exactly one element of $S$.
\end{definition}

\begin{lemma}[\cite{CD06}]\label{spread}
A finite projective space $PG(N,q)$ contains a $t$-spread if and only if $(t + 1) \mid (N+ 1)$. In particular, there exists a spread of lines in $PG(3,q)$ of size $q^2+1$.
\end{lemma}

\begin{definition}[Sunflowers in projective space \cite{ES16}]
Suppose $t \leq s \leq N$. A \emph{sunflower} $SF_{q}(t, s, N)$ is  a set $S$ of $s$-subspaces of $PG(N, q)$ such that they meet at a common $t$-subspace.  The common $t$-subspace is called the \emph{center} of $SF_{q}(t, s, N)$ and the $s$-subspaces are called \emph{petals} of $SF_{q}(t, s, N)$. A sunflower $SF_{q}(t, s, N)$ is called \emph{maximal} if it has largest possible size for fixed $q, t, s, N$.
\end{definition}

When $s=t+1$, we can calculate the size of a maximal sunflower $SF_{q}(t, t+1, N)$
\begin{lemma}\label{sunflower}
The size of  a maximal sunflower $SF_{q}(t, t+1, N)$ is equal to $\frac{q^{N-t}-1}{q-1}$.
\end{lemma}

\begin{IEEEproof}
By using the affine geometry language, it is equivalent to show that for a fixed $(t+1)$-subspace $V$ of $\mathbb{F}_{q}^{N+1}$, the total number $M$ of $(t+2)$-subspaces containing $V$ is equal to $\frac{q^{N-t}-1}{q-1}$. Let $\{\bm{v}_{1}, \bm{v}_{2}, \cdots, \bm{v}_{t+1}\}$ be a basis of $V$, then the number of vectors $\bm{v}_{t+2}\in \mathbb{F}_{q}^{N+1}$ such that $\bm{v}_{1}, \bm{v}_{2}, \cdots, \bm{v}_{t+2}$ span a $(t+2)$-dimensional subspace of $\mathbb{F}_{q}^{N+1}$ is equal to $q^{N+1}-q^{t+1}$. Let $U$ be any $(t+2)$-subspace containing $V$. Then the number of vectors $\bm{v}_{t+2} \in U$ such that $\bm{v}_{1}, \bm{v}_{2}, \cdots, \bm{v}_{t+2}$ form a basis of $U$ is equal to $q^{t+2}-q^{t+1}$. Thus \[M=\frac{q^{N+1}-q^{t+1}}{q^{t+2}-q^{t+1}}
=\frac{q^{N-t}-1}{q-1}.\]
The lemma is proved.
\end{IEEEproof}

\subsection{Binary Constant Weight Codes}
A binary $(n, M, d; w)$  constant weight code is a set of binary vectors of length $n$ with $M$ vectors, such that each vector contains $w$ ones and $n-w$ zeros, and any two distinct vectors differ in at least $d$ positions. The following is the well-known Johnson bound for binary constant weight codes, which is important for our subsequent derivation.

\begin{lemma}(\cite[Restricted Johnson Bound]{MS77})\label{johnson}
Let $C$ be a binary $(n, M, d=2\delta; w)$  constant weight code, then
\[M(w^2-wn+\delta n) \leq \delta n.\]

\end{lemma}

\section{Singleton-optimal LRCs with $d=6$ and $r=3$}

In this section, we consider  Singleton-optimal LRCs with minimum distance $d=6$ and locality $r=3$. Throughout this section, we assume $4 \mid n, \ell=\frac{n}{4}$.

Suppose $C$ is an LRC of length $n$, dimension $k$, minimum distance $d=6$ and locality $r=3$ with disjoint repair groups. From Sec. II.A, we assume that $C$ has a parity-check matrix $H$ as the following form:
\begin{equation}\label{4}
  H=\left(
  \begin{array}{cccc|cccc|c|cccc}
    1 & 1 &  1 & 1 & 0 & 0 &  0 & 0 & \ldots & 0 & 0 & 0 & 0 \\
    0 & 0 & 0 & 0 & 1 & 1 &  1 & 1 & \ldots & 0 & 0 & 0 & 0\\
    \vdots & \vdots & \vdots & \vdots & \vdots & \vdots &  \vdots & \vdots & \ddots & \vdots & \vdots & \vdots & \vdots  \\
    0 & 0 &  0 & 0 & 0 & 0 & 0 & 0 & \ldots & 1 & 1 & 1 &  1  \\
    \hline
    \bm{0} & \bm{u}_{1} &  \bm{v}_{1} & \bm{w}_{1} &\bm{0} & \bm{u}_{2} & \bm{v}_{2} & \bm{w}_{2} & \ldots & \bm{0} & \bm{u}_{\ell} & \bm{v}_{\ell} & \bm{w}_{\ell} \\
  \end{array}
\right),
\end{equation}
where $\bm u_{i},\bm v_{i}, \bm w_{i} \in \mathbb{F}_{q}^{u},$ $i \in [\ell]$, $u=n-k-\ell$.

For convenience, denote $H=(\bm h_{0,1},\bm h_{1,1},\bm h_{2,1},\bm h_{3,1},\cdots, \bm h_{0,\ell},\bm h_{1,\ell},\bm h_{2,\ell},\bm h_{3,\ell})$, where $\bm h_{i,j} \in \mathbb{F}_{q}^{n-k}$.

\begin{proposition}\label{prop1}
The vectors $\bm u_{i},\bm v_{i}, \bm w_{i}$ defined above are linearly independent over $\mathbb{F}_q$.
\end{proposition}

\begin{IEEEproof}
Suppose $a\bm u_{i}+b\bm v_{i}+c\bm w_{i}=0$, for some $a, b, c \in \mathbb{F}_q$. Then from \eqref{4}, we can deduce that $-(a+b+c)\bm{h}_{0,i}+a\bm{h}_{1,i}+b\bm{h}_{2,i}+c\bm{h}_{3,i}=0$. Since the minimum distance $d =6$, we have $a=b=c=0$. Thus $\bm u_{i},\bm v_{i}, \bm w_{i}$ are linearly independent
\end{IEEEproof}

Let
\[L_{1,i}\triangleq \textnormal{Span}_{\mathbb{F}_{q}}\{\bm u_{i},\bm v_{i}\} \subseteq \mathbb{F}_{q}^{u}, \]
\[L_{2,i}\triangleq \textnormal{Span}_{\mathbb{F}_{q}}\{\bm v_{i},\bm w_{i}\} \subseteq \mathbb{F}_{q}^{u},\]
\[L_{3,i}\triangleq \textnormal{Span}_{\mathbb{F}_{q}}\{\bm w_{i},\bm u_{i}\} \subseteq \mathbb{F}_{q}^{u},\]
and
\[L_{4,i}\triangleq \textnormal{Span}_{\mathbb{F}_{q}}\{a\bm u_{i}+b\bm v_{i}+c\bm w_{i}: a,b,c \in \mathbb{F}_{q}, \textnormal{ and }a+b+c=0\} \subseteq \mathbb{F}_{q}^{u}.\]

From Proposition \ref{prop1}, for any $1 \leq i\neq i' \leq 4$ and $1 \leq j \leq \ell$, $\dim(L_{i,j})=2$ and $L_{i,j} \neq L_{i',j}$.

For convenience, we will use the language of projective geometry. Each $L_{i,j}$ can be seen as a line in $PG(u-1,q)$. Moreover, we will use $\bm {u}_{i}$ to represent the point $\textnormal{Span}_{\mathbb{F}_{q}}\{\bm u_{i}\}$ in $PG(u,q)$. In this sense, $\bm u_{i}$ and $\lambda \bm u_{i}$ represent the same point, for any $\lambda \in \mathbb{F}_{q}^{*}$.

\begin{proposition}\label{prop2}
Denote $B_{i}\triangleq \{L_{1,i},L_{2,i},L_{3,i},L_{4,i}\},$ the set of 4 lines $L_{1,i},L_{2,i},L_{3,i},L_{4,i}$ in $PG(u-1,q).$ Then
\begin{description}
      \item[(i)] $B_{i}\bigcap B_{j}=\emptyset$;
      \item[(ii)] For any $1 \leq i \neq j \leq \ell$, there are no 3 lines in $B_{i}\bigcup B_{j}$ through a common point.
    \end{description}
\end{proposition}

\begin{IEEEproof}
(i): $B_{i}\bigcap B_{j}=\emptyset$ follows from  $L_{i,j} \neq L_{i',j}$ for any $1 \leq i\neq i' \leq 4$ and $1 \leq j \leq \ell$.

(ii)
Firstly, there are no 3 lines in $B_{i}$ through a common point. Indeed, the 4 lines in $B_{i}$ produce 6 distinct intersection points: $\bm u_{i}$, $\bm v_{i}$, $\bm w_{i}$, $\bm u_{i}-\bm v_{i}$, $\bm v_{i}-\bm w_{i}$,$\bm w_{i}-\bm u_{i}$.

Then, suppose there are three lines of $B_{i}\bigcup B_{j}$ through a common point. Without loss of generality, we assume that two of them from $B_{i}$ and the other one from $B_{j}$. There are essentially 4 cases need to be considered.
\begin{description}
  \item[1)] Suppose $L_{1,i},L_{2,i}, L_{1,j}$ through a common point. Note that $L_{1,i}$ and $L_{2,i}$ meet at the point $\bm u_{i}$. Thus $\bm u_{i}=a\bm u_{j}+b\bm v_{j}$, for some $a, b \in \mathbb{F}_{q}^{*}$. Then we can verify that $\bm h_{0,i}, \bm h_{1,i}, \bm h_{0,j}, \bm h_{1,j}$ and $\bm h_{2,j}$ are linearly dependent, which leads to $d \leq 5$;
  \item[2)] Suppose $L_{1,i},L_{2,i}, L_{4,j}$ through a common point. We can also verify that $\bm h_{0,i}, \bm h_{1,i}, \bm h_{1,j}, \bm h_{2,j}$ and $\bm h_{3,j}$ are linearly dependent, which leads to $d \leq 5$;
  \item[3)] Suppose $L_{1,i},L_{4,i}, L_{1,j}$ through a common point. Note that $L_{1,i}$ and $L_{4,i}$ meet at the point $\bm{u}_{i}-\bm{v}_{i}$. We can verify that $\bm h_{1,i}, \bm h_{2,i}, \bm h_{0,j}, \bm h_{1,j}$ and $\bm{h}_{2,j}$ are linearly dependent, which also leads to $d \leq 5$.
  \item[4)] Suppose $L_{1,i},L_{4,i}, L_{4,j}$ through a common point.  We can also verify that $\bm h_{1,i}, \bm h_{2,i}, \bm{h}_{1,j}, \bm h_{2,j}$ and $\bm h_{3,j}$ are linearly dependent, which leads to $d \leq 5$.
\end{description}
The proposition is proved.
\end{IEEEproof}

Now, we are already to present a sufficient and necessary condition of the existence of $q$-ary Singleton-optimal LRCs of length $n$, minimum distance $d=6$ and locality $r=3$ with disjoint repair groups.

\begin{theorem}\label{d6r3}
  Suppose $4 \mid n$. Then, there exists a $q$-ary Singleton-optimal LRC of length $n$, minimum distance $d=6$ and locality $r=3$ with disjoint repair groups if and only if
    there exist $\ell$ sets $B_{1}, B_{2}, \ldots, B_{\ell}$, each of which consists of 4 lines in $PG(2,q)$, such that for any $1 \leq i \neq j \leq \ell$,
    \begin{description}
      \item[(i)] $B_{i}\bigcap B_{j}=\emptyset$;
      \item[(ii)] there are no 3 lines in $B_{i}\bigcup B_{j}$ through a common point.
    \end{description}
\end{theorem}

\begin{IEEEproof}
\textbf{Necessity:} Note that from Eq. \eqref{3}, $u=n-k-\ell=d-2-\left\lfloor\frac{d-2}{r+1}\right\rfloor=3$. Then the conclusion follows from Proposition \ref{prop2} by taking $u=3$.
\vskip 2mm \noindent
\textbf{Sufficiency:} Assume that there exist $\ell$ sets $B_{1}, B_{2}, \ldots, B_{\ell}$ satisfying the conditions (i) and (ii). Suppose $B_{i}= \{L_{1,i},L_{2,i},L_{3,i},L_{4,i}\}.$ Then by condition (ii), the 4 lines $L_{1,i},L_{2,i},L_{3,i},L_{4,i}$ produce 6 distinct intersection points. Suppose $\bm{u}_{i}=L_{1,i}\bigcap L_{3,i}$, $\bm{v}_{i}=L_{1,i}\bigcap L_{2,i}$, $\bm{w}_{i}=L_{2,i}\bigcap L_{3,i}$, where $\bm{u}_{i},\bm{v}_{i}, \bm{w}_{i} \in \mathbb{F}_{q}^{3}$. Then $\bm{u}_{i},\bm{v}_{i}$ and  $\bm{w}_{i}$ are linearly independent. Thus
\[L_{1,i}= \textnormal{Span}_{\mathbb{F}_{q}}\{\bm{u}_{i},\bm{v}_{i}\},\]
\[L_{2,i}= \textnormal{Span}_{\mathbb{F}_{q}}\{\bm{v}_{i},\bm{w}_{i}\},\]
\[L_{3,i}= \textnormal{Span}_{\mathbb{F}_{q}}\{\bm{w}_{i},\bm{u}_{i}\}.\]
Suppose $L_{1,i}\bigcap L_{4,i}=\alpha\bm{u}_{i}-\beta\bm{v}_{i}$ and $L_{2,i}\bigcap L_{4,i}=\gamma\bm{v}_{i}-\delta\bm{w}_{i}$, where $\alpha, \beta, \gamma, \delta \in \mathbb{F}_{q}^{*}$. Without loss of generality, we may assume that $\beta=\gamma$. Moreover, by replacing $\bm{u}_{i},\bm{v}_{i}$ and  $\bm{w}_{i}$ by $\alpha\bm{u}_{i},\beta\bm{v}_{i}$ and  $\delta\bm{w}_{i}$, respectively, we may assume that $L_{1,i}\bigcap L_{4,i}=\bm{u}_{i}-\bm{v}_{i}$ and $L_{2,i}\bigcap L_{4,i}=\bm{v}_{i}-\bm{w}_{i}$. Then we have
\[L_{4,i}\triangleq \textnormal{Span}_{\mathbb{F}_{q}}\{a\bm{u}_{i}+b\bm{v}_{i}+c\bm{w}_{i}: a,b,c \in \mathbb{F}_{q}, \textnormal{ and }a+b+c=0\}.\]
Now we let $C$ be the linear code with parity-check $H$ given as (\ref{4}), where $\bm{u}_{i},\bm{v}_{i}$ and  $\bm{w}_{i}$ are given as above. Then $C$ has locality 3 and the dimension $k \geq n-\ell-3=\frac{3}{4}n-3$. By the Singleton-type bound,
\[d \leq n-k+2-\lceil\frac{k}{3}\rceil \leq 6.\]
From condition (ii) and similar discussions of the proof of Proposition \ref{prop2}, we can verify that any 5 columns of $H$ are linearly independent. Thus $d \geq 6$, hence $d=6$ and $C$ is a $q$-ary Singleton optimal LRC of length $n$, minimum distance $d=6$ and locality $r=3$ with disjoint repair groups.

The proof is completed.
\end{IEEEproof}

According to Theorem \ref{d6r3}, we can immediately obtain the first bound on the code length of Singleton-optimal LRCs with $d=6$ and $r=3$.
\begin{theorem}\label{thm2}
  Suppose $4 \mid n$. Let $C$ be a $q$-ary Singleton-optimal LRC of length $n$, $d=6$ and locality  $r=3$. Then
   \[n \leq 4\lfloor\frac{q^{2}-3q+9}{6}\rfloor.\]
\end{theorem}

 \begin{IEEEproof}
 Let $B_i, i=1,2, \cdots, \ell$ the set of lines in $PG(2,q)$ given by Theorem \ref{d6r3}. Note that the 4 lines in the set $B_{i}$ produce 6 distinct intersection points. Let $E_i$ be the set of these 6 points. By condition (ii) of Theorem \ref{d6r3}, all these $6\ell$ intersection points are mutually distinct and each line of $B_{i}$ doest not through any points in $E_j$ ($1\leq j \neq i \leq \ell$), thus $6\ell +4(q+1-3)\leq q^{2}+q+1$, i.e., $\ell \leq \lfloor\frac{q^{2}-3q+9}{6}\rfloor.$ Hence
 \[n =4 \ell \leq 4\lfloor\frac{q^{2}-3q+9}{6}\rfloor.\]
 \end{IEEEproof}

Furthermore, by using the techniques of incidence matrix and Johnson bound on constant weight codes, we can deduce a tighter upper bound.

\begin{theorem}\label{thm4}
  Suppose $4 \mid n$. Let $C$ be a $q$-ary Singleton-optimal LRC of length $n$, minimum distance $d=6$ and locality $r=3$ with disjoint repair groups. Then
  \[n \leq 4\lfloor\frac{7q+3+\sqrt{24q^3+q^2-6q-63}}{24}\rfloor=O(q^{1.5}).\]
\end{theorem}

\begin{IEEEproof}
   Let $B_i,i=1,2, \cdots, \ell$ the set of lines in $PG(2,q)$ given by Theorem \ref{d6r3}. Let $E_i$ be the set of 6 points produced by the 4 lines in $B_i$. According to Theorem \ref{d6r3} (ii), $E_i \cap E_j =\emptyset$. Denote $E=\bigcup_{i=1}^{\ell}E_i$.

   We consider a $4\ell \times (q^{2}+q+1)$ binary matrix $A=(a_{ij})$ whose rows are indexed by the $4\ell$ lines in $B_{1}, B_{2}, \ldots, B_{\ell}$, and columns are indexed by the points of $PG(2,q)$. The $(i, j)$-th entry $a_{ij}=1$ if and only if the ``$i$-th line'' through the ``$j$-th point''. Note that each line in $PG(2,q)$ has $q+1$ points (Lemma \ref{lem2}), thus each row of $A$ has weight $q+1$.

   Let $G$ be the sub-matrix of $A$ consists of the $6\ell$ columns indexed by $E$. According to Theorem \ref{d6r3}, for each line $L_i$ of $B_i$, there are exactly three points of $E_i$ lie on $L_i$ and no points of $E\setminus E_i$ lie on $L_i$. So we can deduce that each row of $G$ has weight 3. Let $M$ be the matrix obtained by deleting $G$ in $A$. Then $M$ is a $4\ell \times (q^{2}+q+1-6\ell)$ binary matrix and each row of $M$ has weight $q+1-3=q-2$. Let $D$ be the binary code of length $q^{2}+q+1-6\ell$ whose codewords are the rows of $M$. Note that any two lines in $PG(2,q)$ meet at exactly one point (Lemma \ref{lem2}). Thus any two distinct codewords of $D$ have Hamming distance at least $2(q-2)-2=2q-6$. So $D$ is a binary $(q^{2}+q+1-6\ell, M=4\ell, d=2q-6; w=q-2)$-constant weight code. By Johnson bound (Lemma \ref{johnson}),
  \[4\ell ((q-2)^{2}-(q^{2}+q+1-6\ell))\leq (q^{2}+q+1-6\ell)(q-3),\]
  \[\Rightarrow 24\ell^{2}-(14q+6)\ell-(q^{2}+q+1)(q-3) \leq 0,\]
  \[\Rightarrow \ell \leq \frac{7q+3+\sqrt{24(q^{2}+q+1)(q-3)+(7q+3)^{2}}}{24}=\frac{7q+3+\sqrt{24q^3+q^2-6q-63}}{24}.\]
  Thus
 \[ n=4\ell \leq 4\lfloor\frac{7q+3+\sqrt{24q^3+q^2-6q-63}}{24}\rfloor=O(q^{1.5}).\]
\end{IEEEproof}

\begin{remark}
From Table II, Theorem \ref{thm4} improves the previous bounds on the code of length of $(n, k, d=6; r=3)$ Singleton-optimal LRCs.
\end{remark}

By the duality of the projective plane (see Theorem \ref{dual}), we present the dual version of Theorem \ref{d6r3} as follows.

\begin{theorem}[Dual version of Theorem \ref{d6r3}]\label{thm'}
  Suppose $4 \mid n$. Then, there exists a $q$-ary Singleton-optimal LRC of length $n$, minimum distance $d=6$ and locality $r=3$ with disjoint repair groups if and only if
    there exist $\ell$ sets $S_{1}, S_{2}, \ldots, S_{\ell}$, each of which consists of 4 points in $PG(2,q)$, such that for any $1 \leq i \neq j \leq \ell$,
    \begin{description}
      \item[(i)] $S_{i}\bigcap S_{j}=\emptyset$;
      \item[(ii)] there are no 3 points in $S_{i}\bigcup S_{j}$ lie on a common line.
    \end{description}
\end{theorem}

If we write $S_i=\{\bm a_{i,1},\bm a_{i,2},\bm a_{i,3},\bm a_{i,4}\}$, where each $\bm a_{i,t} \in \mathbb{F}^3_q$. Then we may restate Theorem \ref{thm'} in the affine geometrical language as follows.

\begin{theorem}[Affine version of Theorem \ref{thm'}]\label{thm''}
  Suppose $4 \mid n$. Then, there exists a $q$-ary Singleton-optimal LRC of length $n$, minimum distance $d=6$ and locality $r=3$ with disjoint repair groups if and only if
    there exist $3 \times n$ matrix  $G=(\bm g_1, \bm g_2, \cdots,\bm g_n),$ where $\bm g_i \in \mathbb{F}^3_q$, such that the columns of $G$ can be divided into $n/4$ groups $G_1, G_2, \cdots, G_{\frac{n}{4}}$ satisfying $|G_i|=4$ and any 3 columns from $G_i$ and $G_j$  are linearly independent.
\end{theorem}

Finally, we will use Theorem \ref{thm''} to construct a family of $q$-ary Singleton-optimal LRCs of $d=6, r=3$ and length $n=q+4$, for $q=2^m$.

Suppose $\mathbb{F}^{*}_q=\{\alpha_1, \alpha_2, \cdots, \alpha_{q-1}=1\}$. Let

\[G=\left(\begin{array}{ccccccccc}
    1& 1& \cdots & 1 &1 &0 & 0 &1 &0 \\
     \alpha_1&  \alpha_2&\cdots & \alpha_{q-1}& 0&1&0&0&1 \\
     \alpha^2_1&  \alpha^2_2&\cdots & \alpha^2_{q-1}& 0&0&1&1&1
\end{array}\right).\]
It is easy to see that any 3 columns of first $q+2$ columns of $G$ are linearly independent.

For $\alpha \in \mathbb{F}^{*}_q$, denote $\bm v(\alpha)=\left(
  \begin{array}{c}
     1\\
     \alpha\\
   \alpha^2 \\
  \end{array}
\right).$

The following two lemmas follows from direct calculations.
\begin{lemma}\label{lem3}
Suppose $\alpha \neq \beta \in \mathbb{F}^{*}_q$
Then
\begin{description}
  \item[(i)] \[ \bm v(\alpha),\bm v(\beta), \left(
  \begin{array}{c}
    1 \\
    0 \\
     0\\
  \end{array}
\right)\] are linearly dependent if and only if $\alpha \beta=1$.
\item[(ii)] \[ \bm v(\alpha),\bm v(\beta), \left(
  \begin{array}{c}
    0 \\
    1 \\
     1\\
  \end{array}
\right)\] are linearly dependent if and only if $\alpha + \beta=1$.
\item[(iii)]
\[ \bm v(\alpha), \left(
  \begin{array}{c}
    1 \\
    0 \\
     1\\
  \end{array}
\right), \left(
  \begin{array}{c}
    0 \\
    1 \\
     1\\
  \end{array}
\right)\] are linearly dependent if and only if $\alpha$ is a cubic primitive root pf unity.

\item[(iv)]
\[ \bm v(\alpha), \left(
  \begin{array}{c}
    1 \\
    0 \\
     1\\
  \end{array}
\right), \left(
  \begin{array}{c}
    0 \\
    1 \\
     0\\
  \end{array}
\right)\] are linearly dependent if and only if $\alpha=1$.
\end{description}
\end{lemma}

\begin{lemma}\label{lem4}
 Suppose $\alpha, \beta \neq 1 \in \mathbb{F}^{*}_q$ are not a  cubic primitive root pf unity. Define
 \[A(\alpha)=\{\alpha, 1+\alpha, \frac{1}{1+\alpha}, \frac{\alpha}{1+\alpha}, \frac{1+\alpha}{\alpha}, \frac{1}{\alpha}\}.\]
 Then
 \begin{description}
   \item[(i)] $|A|=6$
   \item[(ii)] $A+1=\{1+a: a\in A\}=A$
   \item[(iii)] $\frac{1}{A}=\{\frac{1}{a}: a\in A\}=A$
   \item[(iv)] $A(\alpha)=A(\beta)$ if and only if $\alpha=\beta$, or
 $\alpha+\beta=1$, or $\alpha \beta=1$.
 \end{description}
\end{lemma}

Now we present our construction as follows.
\begin{theorem}\label{thm7}
Suppose $q=2^m, m \geq 3$ and $n=q+4$. Then there exists a $q$-ary Singleton-optimal LRC of length $n=q+4$, minimum distance $d=6$ and locality $r=3$.
\end{theorem}

\begin{IEEEproof}
\textbf{Case 1:} $m$ is odd. Then $\mathbb{F}^{*}_q$ does not contain cubic primitive root pf unity.  Choose $\xi \neq 1 \in \mathbb{F}^{*}_q$. Let
\[G_1=\left\{\left(
  \begin{array}{c}
    1 \\
    0 \\
     0\\
  \end{array}
\right), \left(
  \begin{array}{c}
    0 \\
     1\\
     0\\
  \end{array}
\right), \bm v(\xi),\bm v(\frac{\xi}{1+\xi})\right\},\]
\[G_2=\left\{\left(
  \begin{array}{c}
    0 \\
    0 \\
     1\\
  \end{array}
\right), \left(
  \begin{array}{c}
    1 \\
     1\\
     1\\
  \end{array}
\right), \bm v(1+\xi), \bm v(\frac{1+\xi}{\xi})\right\},\]
\[G_3=\left\{\left(
  \begin{array}{c}
    1 \\
    0 \\
     1\\
  \end{array}
\right), \left(
  \begin{array}{c}
    0 \\
     1\\
     1\\
  \end{array}
\right), \bm v(\frac{1}{1+\xi}),\bm v(\frac{1}{\xi})\right\}.\]
From Lemmas \ref{lem3} and \ref{lem4}, we can verify that any 3 column vectors from two groups of $G_1, G_2, G_3$ are linearly independent. For example, the three vectors $\left(
  \begin{array}{c}
    1 \\
    0 \\
     0\\
  \end{array}
\right), \left(
  \begin{array}{c}
    0 \\
     1\\
     0\\
  \end{array}
\right) \in G_1, \bm v(1+\xi) \in G_2$ are linearly independent. Note that
\[\mathbb{F}^{*}_q \setminus (\{1\}\cup A(\xi)) = \bigcup_{\alpha \neq 1, \xi}A(\alpha).\]
Thus we can divided the remain $n-12$ column vectors $\{\bm v(\alpha)\}_{\alpha \notin \{1\}\cup A(\xi)}$ into  $\frac{n-12}{4}$ groups $G_i=\{\bm v(a_i), \bm v(b_i), \bm v(c_i), \bm v(d_i)\}, i=3, 4, \cdots, \frac{n}{4}$, such that $A(a_i), A(b_i), A(c_i), A(d_i)$ are mutually distinct. It is obvious that any 3 column vectors from $G_3, G_4, \cdots, G_{\frac{n}{4}}$ are linearly independent. By Lemmas \ref{lem3} and \ref{lem4} again, we can verify that 3 column vectors from one group of $G_1, G_2, G_3$ and one group of $G_3, G_4, \cdots, G_{\frac{n}{4}}$ are linearly independent. Thus the conclusion follows from Theorem \ref{thm''}.

\textbf{Case 2:} $m \geq 4$ is even. Let $\omega \in \mathbb{F}^{*}_q$ be a cubic primitive root pf unity. Suppose $\alpha, \beta \in \mathbb{F}^{*}_q\setminus \{\omega\}$ with $A(\alpha) \neq A(\beta).$
Let
\[G_1=\left\{\left(
  \begin{array}{c}
    0 \\
    0 \\
     1\\
  \end{array}
\right), \left(
  \begin{array}{c}
    1 \\
     1\\
     1\\
  \end{array}
\right), \bm v(\omega), \bm v(\alpha)\right\},\]
\[G_2=\left\{\left(
  \begin{array}{c}
    1 \\
    0 \\
     0\\
  \end{array}
\right), \bm v(\omega^2),\bm v(1+\alpha),\bm v(\frac{1}{\alpha})\right\},\]

\[G_3=\left\{\left(
  \begin{array}{c}
    0 \\
    1 \\
     0\\
  \end{array}
\right),\bm v(\frac{1}{1+\alpha}), \bm v(\beta),\bm v(\frac{\beta}{1+\beta})\right\},\]

\[G_4=\left\{\left(
  \begin{array}{c}
    1 \\
    0 \\
     1\\
  \end{array}
\right),\bm v(\frac{\alpha}{1+\alpha}), \bm v(1+\beta),\bm v(\frac{1+\beta}{\beta})\right\},\]
\[G_5=\left\{\left(
  \begin{array}{c}
    0 \\
    1 \\
     1\\
  \end{array}
\right),\bm v(\frac{1+\alpha}{\alpha}),\bm v(\frac{1}{1+\beta}),\bm v(\frac{1}{\beta})\right\},\]
From Lemmas \ref{lem3} and \ref{lem4}, we can verify that any 3 columns from two groups of $G_1, G_2, G_3, G_4, G_5$ are linearly independent. The rest of the proof is similarly as the proof of the case of $m$ is odd, whcih we omit it here.

The theorem is proved.
\end{IEEEproof}

\begin{remark}
To the best of our knowledge, Theorem \ref{thm7} is the first construction of $q$-ary Singleton-optimal LRCs of minimum distance $d=6$, locality $r=3$ and length $n >q+1$.
\end{remark}

\section{Singleton-optimal LRCs with $d=7$ and $r=2$}
In this section, we consider  Singleton-optimal LRCs with minimum distance $d=7$ and locality $r=2$. Throughout this section, we assume
 $3 \mid n, \ell=\frac{n}{3}$.

 Suppose $C$ is an LRC of length $n$, dimension $k$, minimum distance $d=7$ and locality $r=2$ with disjoint repair groups. From Sec. II.A, we assume that $C$ has a parity-check matrix $H$ as the following form:
\begin{equation}\label{5}
  H=\left(
  \begin{array}{ccc|ccc|c|ccc}
    1 &   1 & 1 & 0 &   0 & 0 & \ldots & 0  & 0 & 0 \\
    0  & 0 & 0 & 1  &  1 & 1 & \ldots & 0  & 0 & 0\\
    \vdots  & \vdots & \vdots & \vdots &   \vdots &  \ddots & \vdots & \vdots & \vdots & \vdots  \\
    0 &   0 & 0 & 0 & 0  & 0 & \ldots & 1 &  1 &  1  \\
    \hline
    \bm{0} & \bm{u}_{1} &  \bm{v}_{1} & \bm{0} & \bm{u}_{2} & \bm{v}_{2} &  \ldots & \bm{0} & \bm{u}_{\ell} & \bm{v}_{\ell}  \\
  \end{array}
\right),
\end{equation}
where $\bm u_{i},\bm v_{i} \in \mathbb{F}_{q}^{u},$ $i \in [\ell]$, $u=n-k-\ell$.

We present the sufficient and necessary condition of the existence of $q$-ary Singleton-optimal LRCs of length $n$, minimum distance $d=7$ and locality $r=2$ with disjoint repair groups as follows.
\begin{theorem}\label{thm8}
  Suppose $3 \mid n$. Then, there exists a $q$-ary Singleton-optimal LRC of length $n$, minimum distance $d=7$ and locality $r=2$ with disjoint repair groups if and only if
    there exist $\ell$ Lines $L_{1}, L_{2}, \ldots, L_{\ell}$  in $PG(3,q)$, such that
    \begin{description}
      \item[(i)] for any $1 \leq i \neq j \leq \ell$, $L_{i}\bigcap L_{j}=\emptyset$;
      \item[(ii)] there exist three distinct points $P_{i,1}, P_{i,2}, P_{i,3}$ in $L_i$ ($i=1,2, \cdots, \ell$) satisfying that $P_{s, \mu}, P_{t, \nu}, P_{m, \omega}$ do not lie on a common line for any $1 \leq s < t < m \leq \ell$ and $\mu, \nu, \omega \in \{1,2,3\}$.
    \end{description}
\end{theorem}

\begin{IEEEproof}
   \textbf{Necessity:} Suppose $C$ is a $q$-ary Singleton-optimal LRC with length $n$, minimum distance $d=7$ and locality $r=2$. Let $H$ be a parity-check matrix of $C$ given by Eq. \eqref{5}. From Remark \ref{rem1} and Eq. \eqref{3}, we have $u=4$.
   Let \[L_{i}\triangleq \textnormal{Span}_{\mathbb{F}_{q}}\{\bm u_{i},\bm v_{i}\} \subseteq \mathbb{F}^4_{q}.\]
   Similar to Proposition \ref{prop1}, we can show that $\dim(L_i)=2$. Thus we can regard $L_i$ as a line in $PG(3,q).$ Suppose $\bm x \in L_{i}\bigcap L_{j}$, then $\bm x=a_i\bm u_i+b_i \bm v_i=c_j\bm u_j+ d_j \bm v_j$. If we denote the columns of $H$ as $\bm h_{1,0}, \bm h_{1,1},\bm h_{1,2}, \cdots, \bm h_{\ell,0},\bm h_{\ell,1},\bm h_{\ell,2},$ then $(a_i+b_i)\bm h_{i,0}-a_i\bm h_{i,1}-b_i \bm h_{i,2}=(c_j+d_j)\bm h_{j,0}-c_j\bm h_{j,1}-d_j \bm h_{j,2}$. Since $d \geq 7,$ we have $a_i=b_i=c_j=d_j=0$. Thus $L_{i}\bigcap L_{j}=\emptyset$. Denote $P_{i,1}, P_{i,2}, P_{i,3}$ the points of $L_i$ represented by $\bm u_{i},\bm v_{i}, \bm u_{i}-\bm v_{i}.$

   We will prove that these points satisfy the condition (ii). By contradiction. W.l.o.g., we suppose $P_{1,1} \in L_1, P_{2,1} \in L_2, P_{3,1} \in L_3$ lie on a common line. Then there exist $a, b, c \in \mathbb{F}^{*}_q$, such that $a\bm u_1+b\bm u_2+c \bm u_3=0$. Then $a(\bm h_{1,1}-\bm h_{1,0})+b(\bm h_{2,1}-\bm h_{2,0})+c(\bm h_{2,1}-\bm h_{2,0})=0,$ which leads to $d \leq 6$. The necessity is proved.
\vskip 2mm \noindent
\textbf{Sufficiency:} Suppose there exist $\ell$ lines $L_{1}, L_{2}, \ldots, L_{\ell}$  in $PG(3,q)$ satisfying the conditions (i) and (ii). Assume that the three points $P_{i,1}, P_{i,2}, P_{i,3}$ in $L_i$ corresponding to the nonzero vectors $\bm u_i, \bm v_i, \bm w_i \in \mathbb{F}^4_{q}.$ Thus $\bm w_i=a\bm u_i-b\bm v_i$, for some $a, b \in \mathbb{F}_{q}^{*}$. By replacing $\bm u_{i}$ and $\bm v_{i}$ with $a\bm u_{i}$ and $b\bm v_{i}$, respectively, we can assume that $\bm w_{i}=\bm u_{i}-\bm v_{i}$. Now we let $C$ be the linear code with parity-check $H$ given as Eq. \eqref{5}. Obviously, $C$ is a $q$-ary LRC of length $n=3\ell$, locality $r=2$, dimension $k\leq n-\ell-4=2\ell-4$. From Lemma \ref{lem1} and Remark \ref{rem1}, we know that $d\leq n-k+1-\lceil\frac{k}{r}\rceil \leq 7$. Thus we only need to prove that $d \geq 7$. By contradiction, suppose there are 6 columns of $H$ are linearly dependent. We divide our discussions into 3 cases.

\textbf{Case 1:} These 6 columns from two repair groups. W.l.o.g., we suppose $\bm h_{1,0}, \bm h_{1,1}, \bm h_{1,2}, \bm h_{2,0}, \bm h_{2,1},\bm h_{2,2}$ are linearly dependent. Then it can deduce that $a\bm u_1+b \bm v_1+c\bm u_2+ d \bm v_2=0$ for some  $a, b, c, d \in\mathbb{F}_q$. Note that $a\bm u_1+b \bm v_1 \in L_1$  and $c\bm u_2+ d \bm v_2 \in L_2$, which leads to $L_i \cap L_j \neq \emptyset.$

\textbf{Case 2:} These 6 columns from three repair groups, and each group contains exactly two of these columns. W.l.o.g., we suppose $\bm h_{1,1}, \bm h_{1,2}, \bm h_{2,1}, \bm h_{2,2}, \bm h_{3,1},\bm h_{3,2}$ are linearly dependent. Then $\bm u_1-\bm v_1, \bm u_2-\bm v_2, \bm u_3-\bm v_3$ are linearly dependent, hence $P_{1,3}, P_{2,3}, P_{3,3}$ lie on a common line, which contradicts with condition (ii).

\textbf{Case 3:} There is a repair group that contains exactly one of these 6 columns. Then this column can not be a linear combination of other columns. Thus these 6 columns are linearly independent.

The sufficiency is proved and the theorem follows.
\end{IEEEproof}

From Theorem \ref{thm8}, we can derive a bound on the code length of $q$-ary $(n, k, d=7; r=2)$ Singleton-optimal LRCs with disjoint repair groups.

\begin{theorem}\label{thm9}
  Suppose $3 \mid n$. Let $C$ be a $q$-ary Singleton-optimal LRC of length $n$, minimum distance $d=7$ and locality $r=2$ with disjoint repair groups. Then
  \[n \leq 3\lfloor\frac{q^2+q+3}{3}\rfloor.\]
\end{theorem}

\begin{IEEEproof}
   Let  $L_{1}, L_{2}, \ldots, L_{\ell}$ be $\ell$ lines in $PG(q,3)$ satisfying the conditions (i) and (ii) in Theorem \ref{thm8}. Consider the $3(\ell-1)+1$ lines $\overline{P_{1,1}P_{i,\mu}}$ ($i=2, 3, \cdots, \ell, \mu=1,2,3)$ and $L_1$. By conditions (i) and (ii) in Theorem \ref{thm8}, we know that these $3(\ell-1)+1$ lines are mutually distinct. Note that the number of lines in $PG(3,q)$ through a fixed point is $\frac{q^3-1}{q-1}=q^2+q+1$ (see Lemma \ref{lem2}). Thus
   \[3(\ell-1)+1 \leq q^2+q+1,\]
   hence $\ell \leq \lfloor\frac{q^2+q+3}{3}\rfloor,$ and
   \[n =3\ell \leq 3\lfloor\frac{q^2+q+3}{3}\rfloor.\]
\end{IEEEproof}

\begin{remark}
From Table II, Theorem \ref{thm9} improves the previous bounds on the code length of $q$-ary $(n, k, d=7; r=2)$ Singleton-optimal LRCs.
\end{remark}
In the following, we use the spread of lines in $PG(3,q)$ to show the existence of $q$-ary Singleton-optimal LRCs with length $n=3\lceil\frac{7+\sqrt{72q^2+121}}{18}\rceil$, minimum distance $d=7$ and locality $r=2$.
\begin{theorem}\label{thm10}
 For any $q\geq 3$, there exists a $q$-ary Singleton-optimal LRC with length $n=3\lceil\frac{7+\sqrt{72q^2+121}}{18}\rceil$, minimum distance 7 and locality 2.
\end{theorem}

\begin{IEEEproof}
  By Lemma \ref{spread}, suppose $S$ is a spread of lines in $PG(3,q)$, that is $S=\{L_1, L_2, \cdots, L_{q^2+1}\}$, where $L_i$ is a line of $PG(3,q)$ with $L_i \cap L_j= \emptyset,$ for any $1 \leq i \neq j \leq q^2+1$. Suppose $m$ is the maximal number of lines in $S$ satisfying the condition (ii) of Theorem \ref{thm8}. Obviously, $m \geq 2$. W.l.o.g, we assume $L_1, L_2, \cdots, L_m$ satisfy the condition (ii) of Theorem \ref{thm8}. By condition (ii), we can show that the lines $\overline{P_{i,\mu}P_{j, \nu}}$, $(1 \leq i <j \leq m, \mu, \nu \in \{1,2,3\})$ are mutually distinct, thus the number of these lines is $3m(3m-3)/2=\frac{9m(m-1)}{2}$. Due to the maximality of $m$, for each $\tau=m+1, \cdots, q^2+1$, the number of points on $L_{\tau}$ which also lie on $\overline{P_{i,\mu}P_{j, \nu}}$ for some $1 \leq i <j \leq m$ and $\mu, \nu \in \{1,2,3\}$ is no less than $q-1$, i.e., for any $m+1 \leq \tau \leq q^2+1$
  \begin{equation}\label{6}
      |\{\bigcup_{1 \leq i <j \leq m\textnormal{~and~} \mu, \nu \in \{1,2,3\}}(L_{\tau} \cap \overline{P_{i,\mu}P_{j, \nu}})\}|\geq q-1.
  \end{equation}
 Otherwise, there exist three points $P_{\tau,1}, P_{\tau, 2}, P_{\tau, 3}$ on $L_{\tau}$ which do not lie on any lines $\overline{P_{i,\mu}P_{j, \nu}}$. Then we obtain $m+1$ lines $L_1, L_2, \cdots, L_m, L_{\tau}$ satisfy the condition (ii) of Theorem \ref{thm8}.

On the other hand,  $P_{i,\mu}, P_{j, \nu} \in  \overline{P_{i,\mu}P_{j, \nu}} \cap \bigcup^{m}_{s=1} L_s$ ($1 \leq i <j \leq m$ and $\mu, \nu \in \{1,2,3\}$). Since $L_s \cap L_{\tau}=\emptyset$ ($s=1,2,\cdots, m$ and $\tau=m+1, \cdots, q^2+1$), from Eq. \eqref{6}, we have
  \[(q+1-2)\frac{9m(m-1)}{2} \geq (q-1)(q^2+1-m),\]
  i.e., $9m^2-7m-2(q^2+1) \geq 0, $ which leads to $m \geq \lceil\frac{7+\sqrt{72q^2+121}}{18}\rceil$.
  Thus there exist at least $\lceil\frac{7+\sqrt{72q^2+121}}{18}\rceil$ lines satisfy the conditions (i) and (ii) of Theorem \ref{thm8}.  The theorem then follows.

\end{IEEEproof}

Finally, we give another construction of $q$-ary Singleton-optimal LRCs with length $n=3\lceil\frac{7+\sqrt{8q^2-16q-7}}{6}\rceil$, minimum distance $d=7$ and locality $r=2$ by the sunflowers in $PG(3,q)$.

\begin{theorem}\label{thm11}
  For any $q\geq 3$, there exists a $q$-ary Singleton-optimal LRC with length $n=3\lceil\frac{7+\sqrt{8q^2-16q-7}}{6}\rceil$, minimum distance 7 and locality 2.
\end{theorem}

\begin{IEEEproof}
  Fix a line $L$ of $PG(3,q)$ and suppose the points of $L$ are $A_1, A_2, \cdots, A_{q+1}.$ Consider the sunflower $SF(1,2,3)$ in $PG(3,q)$ with center $L$. Each petal $\pi_i$ of $SF(1,2,3)$ is a plane through $L$. Let $B_i( \neq A_i)$ be any point in $\pi_i$, it is easy to show that $\overline{A_i B_i}\cap \overline{A_j B_j}=\emptyset,$ for any $1 \leq i \neq j \leq q+1.$ Now, let $L_1$ and $L_2$ be two lines of $\pi_1, \pi_2$ which through $P_{1, 1}=A_1$ and $P_{2,1}=A_2$, respectively. Choose any points $P_{1,2} \neq P_{1,3} \in L_1 \setminus \{P_{1,1}\}$ and $P_{2,2}\neq P_{2,3}\in L_2 \setminus \{P_{2,1}\}$. Suppose $m$ is the maximal number such that we can choose $m$ lines $L_i \subseteq \pi_i$ with $A_i \in L_i$ ($i=1,2, \cdots, m$), and satisfying the condition (ii) of Theorem \ref{thm8}. We consider the lines $\overline{P_{i,\mu}P_{j, \nu}}$, $(1 \leq i <j \leq m, \mu, \nu \in \{1,2,3\})$. Then except for the line $\overline{P_{1, 1}P_{2, 1}}(=L$), the other lines do not lie on the plane $\pi_{m+1}$, hence each of these lines meet $\pi_{m+1}$ at exactly one point.
  Note that among these lines, there are $6(m-1)-1=6m-7$ lines through $P_{1,1}$ or $P_{2,1}$. Now we consider the $q+1$ lines $l_1=L, l_2, \cdots, l_{q+1}$ in $\pi_{m+1}$ which through the common point $A_{m+1}$. If $\frac{9m(m-1)}{2}-(6m-7) < q(q+1-3),$ then we can find three points $A, B, C$ in $l_{j}\setminus\{A_{m+1}\}$ for some $2 \leq j\leq q+1$ such that they do not lie on any lines $\overline{P_{i,\mu}P_{j, \nu}}$, $(1 \leq i <j \leq m, \mu, \nu \in \{1,2,3\})$.  Then the $m+1$ lines $L_1, \cdots, L_m, l_j$ satisfy the conditions of Theorem \ref{thm8}, which contradicts with the maximality of $m$. Thus
  \[\frac{9m(m-1)}{2}-(6m-7) \geq q(q+1-3),\]
  \[\Rightarrow 9m^2-21m-2(q^2-2q-7) \geq 0,\]
  \[\Rightarrow m\geq \lceil\frac{7+\sqrt{8q^2-16q-7}}{6}\rceil.\]
  Thus there exist at least $\lceil\frac{7+\sqrt{8q^2-16q-7}}{6}\rceil$ lines satisfy the conditions (i) and (ii) of Theorem \ref{thm8}. The theorem then follows.
  \end{IEEEproof}
  \begin{remark}
  Both of Theorems \ref{thm10} and \ref{thm11} show the existence of $q$-ary $(n, k, d=7;r=2$ Singleton-optimal LRCs with $n \approx \sqrt{2}q$.
  \end{remark}
\section{Conclusion}
In this paper, we investigate $(n, k, d; r)$ Singleton-optimal LRCs with disjoint repair groups for $d=6, r=3$ and $d=7, r=2$, respectively. We reduce the existence of these optimal LRCs to the existence of some lines in the projective space with certain properties. By some techniques of finite field and finite geometry, some new constructions and bounds of Singleton-optimal LRCs. To the best of our knowledge, our new LRCs have achieved
longer code length and our new bounds on the code length are tighter than previous known results. It is interesting to give tighter upper bounds and some explicit constructions with larger code length in the future work.

\emph{Note added.} --Near the completion of the manuscript, we become aware of a similar results of $(n, k, d=7; r=2)$ Singleton-optimal LRCs, obtained independently in the recent paper, Ref. \cite{GY22}.

\end{document}